# Selective Tweezing and Immobilization of Colloids for Dexterous Manipulation of Biological Materials

Krishangi Krishna,[1†] Jieliyue Sun,[1†] Zhaowei Jiang,[2] Alec Mccall,[2] Anita Shukla,[2] Kimani C. Toussaint Jr.[1,3†]

[1] PROBE Lab, School of Engineering, Brown University, Providence, RI 02912, USA.
[2] Shukla Lab, School of Engineering, Brown University, Providence, RI 02912, USA.
[3] Brown University Center for Digital Health, Providence, RI 02903, USA.
[*]kimani_toussaint@brown.edu
[†]These authors contributed equally

## Abstract

The assembly of arbitrary 3D structures using nano- to micron-scale colloidal building blocks has broad applications in photonics, electronics, and biology. Combining optical tweezers (OT) with two-photon polymerization (TPP) enables 3D selective tweezing and immobilization of colloids (STIC) without requiring specialized particle functionalization. Unlike traditional approaches, we demonstrate that high-repetition-rate femtosecond laser pulses, rather than continuous-wave lasers, allow optical tweezing at intensities below the TPP threshold. This dual functionality enables both OT and TPP using a single laser source. This platform was applied on S. aureus cells into desired configurations, highlighting its potential for advanced cell patterning. TPP was further used to fabricate intricate 3D microstructures, including microgrooves and cylindrical constructs, facilitating spatially resolved studies of single-cell dynamics and interactions. This work highlights the potential of STIC as a versatile tool for advanced biological applications, including tissue engineering and microbial research.

## Introduction

Manipulation of biological materials at micrometer and submicron-scale has become a critical issue for a wide range of biomedical applications, including tissue engineering, drug development and biosensing. One key example of this is cell patterning, a technique that regulates the spatial arrangement of cells on a substrate surface. This method is pivotal for investigating cellular behaviors, such as adhesion, migration, proliferation, survival, and programmed cell death (apoptosis) under different conditions, thereby providing valuable insights into cell biology [1]. Recent advances in micro- and nanotechnologies, and several manipulation techniques have been developed to manipulate biological materials with high precision while minimizing damage. Microcontact printing, microfluidic devices and dielectrophoresis are widely employed for capturing, sorting and transporting cells, and biomolecules. However, these techniques face challenges in providing sustained control over dynamic processes in complex biological systems. Furthermore, these techniques are primarily confined to 2D *in vitro* models, making them less suitable for replicating more physiologically relevant microenvironments.

Optical tweezers (OT) offers an alternative as it is a non-invasive technique capable of manipulating biological matter in 3D with sub-micrometer precision and piconewton-scale forces in real time. This manipulation can be performed either directly on the target or indirectly by tethering it to a trapped microsphere. OT enables the creation of controlled microenvironments, and has been extensively employed to arrange cells in defined patterns for studying cell-cell interactions, tissue formation, and signaling pathways. The operation of OT relies on the equilibrium between electromagnetic gradient forces and scattering forces, a phenomenon resulting from the radiation pressure exerted by light. In biological applications, a standard OT setup typically employs an infrared continuous-wave (CW) laser with intensities on the order of $MW/cm^2$. However, the minimal threshold laser intensities required for effective trapping can induce thermal effects that pose challenges. These include localized heating, which can lead to thermal damage to biological specimens, and the generation of reactive oxygen species (ROS), which may adversely affect cell propagation, motility, and the expression of stress-response genes. We have previously demonstrated that femtosecond (fs) laser-assisted selective tweezing with ultra-low power (FLASH-UP) significantly outperforms conventional continuous-wave optical tweezers (CW-OT) in trap strength and stiffness, achieving a fivefold improvement for micron-sized particles. This advancement enables the direct manipulation of dielectric particles and biological cells without inducing adverse effects [2].

Additionally, utilizing an ultrafast laser source offers the potential to integrate two-photon polymerization (TPP), an advanced 3D printing technique capable of fast prototyping arbitrary constructs with submicron resolution. By exploiting multiphoton absorption, TPP induces highly localized polymerization at the focal volume of the laser, resulting in the solidification of a precisely defined region within the photosensitive resin. This approach has been widely adopted in the fabrication of cell scaffolds and organ-on-a-chip devices owing to its ability to work with an extensive selection of biocompatible materials. Incorporating TPP into biomaterials manipulation not only allows for particle immobilization at designated sites but also creates environments conducive to sustained biological dynamic activity by providing mechanical support or functional regulation for cells and other biomolecules.

Building on this foundation, we developed a single-laser integrated platform for the selective tweezing and immobilization of colloids (STIC). This platform facilitates the assembly and fabrication of microspheres and transition metal dichalcogenides into customizable 2D and 3D patterns [3, 4]. The STIC system exploits the power-threshold difference between TPP and OT at a central wavelength of 800 nm, enabling precise control of both processes within a single setup. Additionally, the platform incorporates capabilities for multiphoton microscopy, allowing for high-resolution inspection of fabricated structures. In this study, we extend the application of STIC to bacterial cells, specifically *Staphylococcus aureus*, to demonstrate its utility in cell patterning and cell sorting. By leveraging the unique properties of fs laser technology, this work highlights the potential of STIC as a versatile tool for advanced biological applications, including tissue engineering and microbial research.

## Materials and Methods

**Cell Preparation**

*S. aureus* ATCC 25923 was obtained from the American Type Culture Collection (ATCC). Sterile Lysogeny Broth (LB) (Sigma Aldrich, St. Louis, MO) was prepared according to the manufacturer's protocol, containing 10 g/L tryptone, 10 g/L sodium chloride (NaCl), and 5 g/L yeast extract. The bacterial strain was stored in 25% (v/v) glycerol in LB at -80°C until use. A sterile inoculating loop was employed to streak the strain onto a nutrient broth agar plate. A single colony was used to inoculate the culture, which was incubated at 37°C with orbital shaking at 200 rpm for 3 hours. The bacterial cultures were subsequently diluted to an optical density of 0.01 at 600 nm (OD600) and washed twice by centrifugation at 4,000 × g for 10 minutes with 1× phosphate-buffered saline (PBS). The resulting cell pellet was resuspended in 1× PBS.

**Sample Preparation**

The photoresist consists of 8% w/v methacrylated gelatin (GelMA, Advanced BioMatrix) in phosphate buffered saline (PBS) serving as the prepolymer and 4.1 mM Rose Bengal (Sigma Aldrich) acting as the photoinitiator. The experimental solution was then prepared by vortexing 32 μL of the photoresist with 25 μL of *S. aureus* cell solution. Subsequently, 25 μL of the final mixture was drop-cast onto a coverslip and enclosed with a gasket (Thermo Fisher) to prevent sample dehydration.

**Optical Setup**

A tunable femtosecond laser (InSight X3, Spectra Physics) with a pulse width of 120 fs, 80-MHz repetition rate, and a central wavelength of 800 nm was employed for optical trapping (OT) and two-photon polymerization (TPP). The laser beam was spatially filtered using a 75-μm precision pinhole to produce a Gaussian intensity profile and subsequently expanded to a beam diameter of 3 mm. The laser's polarization was aligned horizontally relative to the optical table using a linear polarizer, while the average power at the sample stage was modulated with a half-wave plate. Laser positioning at the sample plane was achieved through a two-dimensional galvanometer beam scanner controlled by LabVIEW software. For brightfield imaging, a 150-W broadband light source was focused onto the sample using a 20×/0.5 NA condenser lens. Backscattered light was passed through a short-pass filter (Thorlabs, FESH0700) to block the laser and was subsequently focused onto an sCMOS camera using a tube lens. For the experimental schematic, readers are referred to our previously published work [3].

**STIC protocol**

A cell was first optically trapped 1-2 μm above the substrate coverslip and manipulated to a desired location within the polymer solution. The position of the cell was immobilized into place by periodically moving the objective lens up and down inducing polymerization by increasing the average power; thereby arresting cell Brownian motion. This process was repeated till the desired pattern was fabricated. The precise control over cell movement and immobilization afforded by this technique is a significant advantage, facilitating the creation of intricate arrangements and structures with exceptional accuracy.

**Results and Discussion**

STIC was initially utilized for the manipulation of bacterial cells to form a 3×3 pattern. In PBS, cells were effectively tweezed at an average power of 10 mW, and with an average speed of 2.5 μm/s. However, when the same approach was applied to cells suspended within the photoresist medium, a significantly higher average optical power was required to achieve comparable levels of manipulation and control. This increase in power demand can be attributed to the elevated viscosity of the photoresist relative to PBS.

Viscosity plays a critical role in OT, as it directly influences the drag force acting on the trapped particles. In higher viscosity environments, particles experience greater resistance to movement, necessitating an increase in trapping power to maintain effective manipulation. This observation underscores the importance of accounting for the rheological properties of the surrounding medium when optimizing laser parameters for precise control of particle or cell behavior. These findings highlight STIC's adaptability in diverse media and its potential for applications requiring precise cellular patterning in complex environments.

Figure 1 illustrates the assembly of *S. aureus* (highlighted in the orange circles), with an average diameter of 0.75 μm to 1 μm, arranged in a 3×3 grid with an intercellular spacing of approximately 1 μm. This assembly was achieved at an average optical power of 9.35 mW, enabling manipulation at an average translation speed of 1.75 μm/s. This distance prevents direct contact between cells, but falls within the range for quorum sensing (cell-cell communication) of *S. aureus* cells within biofilm clusters [5].

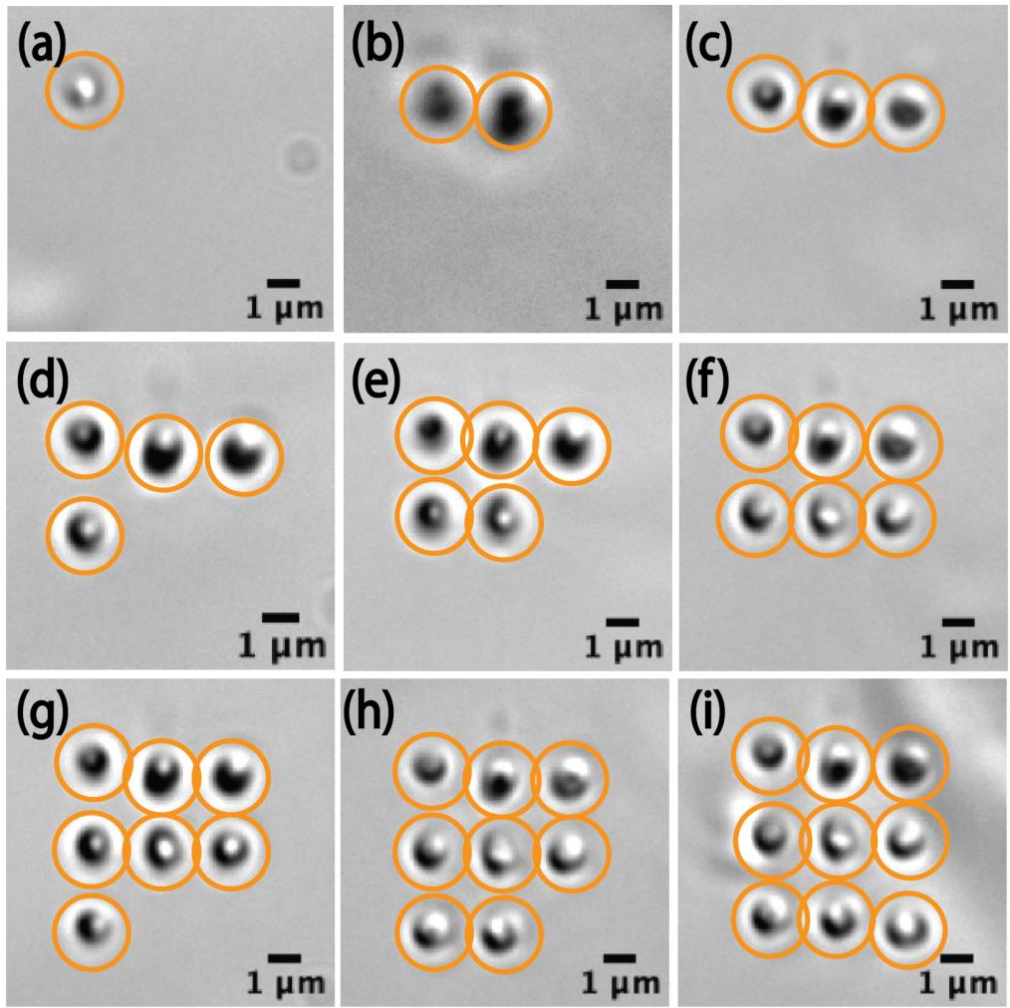

**Fig. 1** Time lapse demonstrating cell patterning of *S. aureus* in a 3×3 grid

Figure 2 demonstrates the integration of microstructures fabricated by TPP in the STIC process. These structures resemble parallel microgroove patterns commonly used in exploring contact guidance in cell engineering. Each pillar had a lateral size of 12.2 μm × 1.2 μm, and consisted of 3 layers spaced evenly at 400 nm axially, as shown in Fig. 2(a). Subsequently, cells were relocated along the walls of the microgrooves using STIC [Figs. 2(b-f)]. No additional polymerization was required for the cells to be securely attached to the wall. As a demonstration, 5 individual cells were immobilized on different pillars.

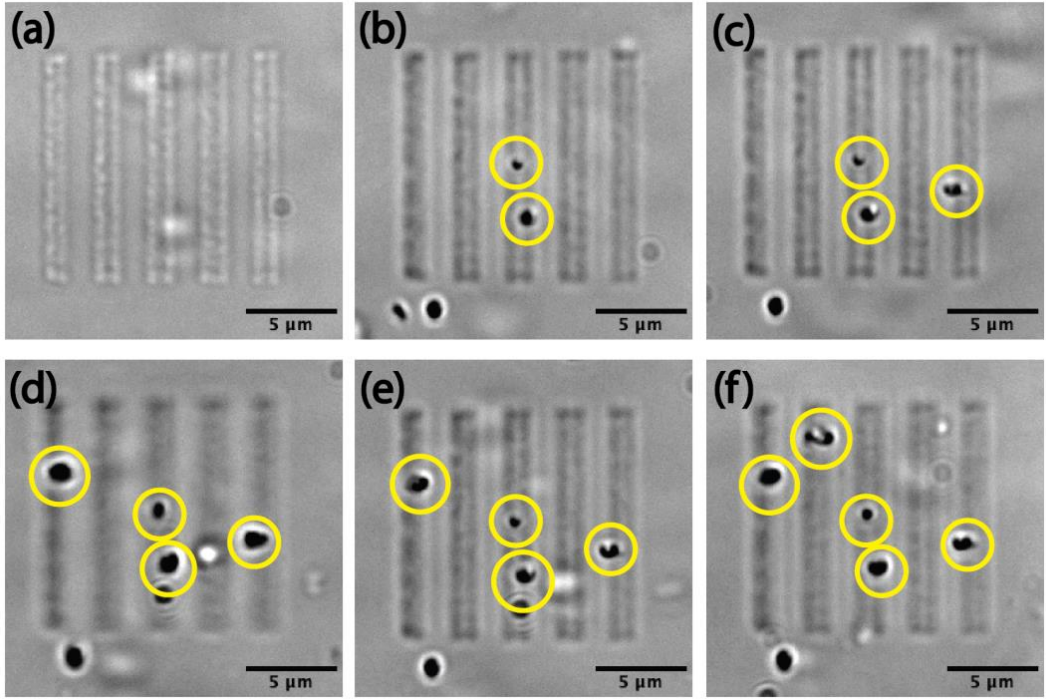

**Fig. 2** (a) Parallel microgroove patterns fabricated by TPP. (b-f) Time lapse images demonstrating the STIC process of *S. aureus* on the wall of the microgrooves

Figure 3 represents a proof-of-concept exploration of single-cell dynamics using STIC. A 2-μm dielectric bead was first immobilized on the substrate, acting as a cell of interest [Fig. 3(a)]. A hollow cylinder ($d$ = 8.5 μm) with ten 1-μm openings was fabricated concentrically around the bead [Fig. 3(b)]. *S. aureus* cells were trapped and relocated to be brought near the bead of interest [Figs. 3 (c-f)].

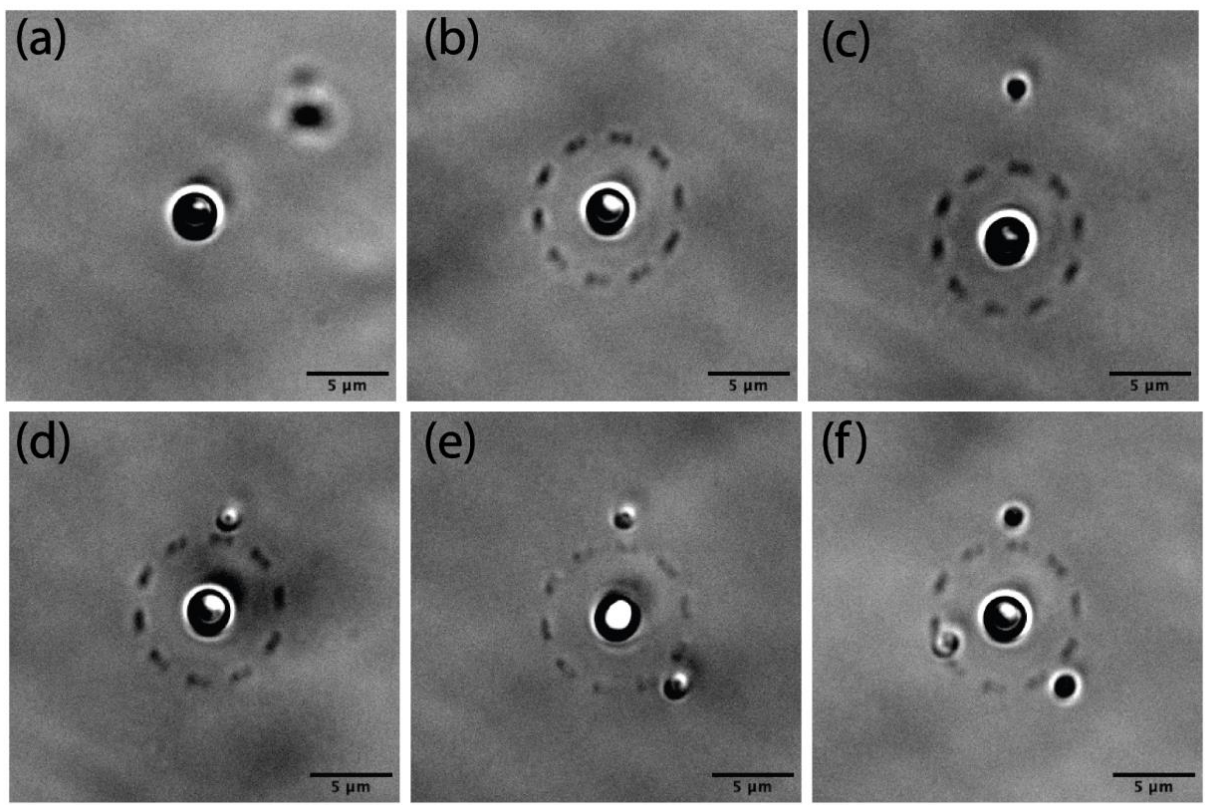

**Fig 3.** A 2-μm polystyrene bead surrounded by a TPP-fabricated hollow cylinder featuring 1-μm openings, with bacteria cells placed arbitrarily on its exterior.

This work establishes a platform capable of mimicking complex biological processes, including drug transport mechanisms across cell layers, leaky vasculature observed in infected or inflamed tissues, and polymicrobial interactions within biofilms. The ability to fabricate customizable microenvironments with sub-micrometer precision further enhances the platform's applicability in studying intercellular signaling, pathogen-host dynamics, and antibiotic resistance mechanisms in a biologically relevant context. This study lays the groundwork for future advancements in microbiology, biophysics, and tissue engineering by providing an adaptable and powerful tool for exploring cell patterning, single-cell encapsulations, and their implications for broader biological systems.

**Conclusion**

In this study, a versatile single-laser platform was developed to enable the precise, contact-free manipulation and assembly of bacterial cells, offering a significant advancement in the field of cell patterning. The system utilizes an ultrafast pulsed laser operating at a central wavelength of 800 nm, which provides the dual capabilities of OT and TPP. *S. aureus* cells were successfully trapped, manipulated, and immobilized into user-defined spatial configurations, demonstrating the system's potential for advanced cell patterning applications. Furthermore, TPP was employed to fabricate intricate 3D microstructures, such as microgrooves and cylindrical constructs to facilitate the spatially resolved study of single-cell dynamics and interactions. These fabricated environments enable controlled studies of microbial behavior, simulating physiologically relevant conditions.

**Acknowledgements**

K.K. and J.S. contributed equally to this work.

**Author Contributions**

K.K. and J.S. designed and conducted the experiments. K.K. and J.S. analyzed the results. Z. J., A. M. provided samples. K.K., J.S., Z.J., A.M., A S., and K.C.T. contributed to the writing and discussion of the manuscript. A.S. and K.C.T. supervised the research

**Funding**

This work was supported by Brown University.

**Data Availability Statement**

The data in this manuscript supporting the findings of this study are available within the paper.

**Declarations**

**Conflict of Interest Statement** The authors declare no conflict of interest.